\documentclass[fleqn,twoside]{article}
\usepackage{espcrc2}
\usepackage{graphicx}
\usepackage[figuresright]{rotating}
\def\Journal#1#2#3#4{{#1} {\bf #2}, #3 (#4)}
\def\NIM{\em Nucl. Instrum. Methods}

\def\PLB{{\em Phys. Lett.}  B}

\newcommand{\AmS}{{\protect\the\textfont2
  A\kern-.1667em\lower.5ex\hbox{M}\kern-.125emS}}

\title{Final $R$-value results from 2-5 GeV from BES and QCD test with 
$R$ scan data}

\author{
W.B. Yan, \address[IHEP]{Institute of High Energy Physics, 
Chinese Academy of Sciences, Beijing 100039, P.R.C.} 
W.G. Li, \addressmark[IHEP]
Z.G. Zhao\addressmark[IHEP]
\address{University of Michigan, Ann Arbor, MI48109, USA}
, Representing BES Collaboration}

\begin{document}

\begin{abstract}

Final results of the measurement of $R = \sigma
(e^+e^- \rightarrow hadrons)/\sigma(e^+e^- \rightarrow \mu^+
\mu^-)$ in the energy region from 2 to 5 GeV with the upgraded
Beijing Spectrometer (BESII) at the Beijing Electron Positron Collider
(BEPC) are presented. 
Preliminary results of the inclusive momentum spectra and second 
binomial moment measured with the $R$ scan data at 2.2, 2.6, 3.0, 3.2, 4.6 
and 4.8 GeV are reported. 

%\vspace{1pc}
\end{abstract}

\maketitle

\section{Introduction}

The lowest order cross section for
$e^+e^-\rightarrow\gamma^*\rightarrow \mbox{hadrons}$
in units of the lowest-order QED cross section for
$e^+e^- \rightarrow \mu^+\mu^-$ is defined as $R$,
namely $R=\sigma(e^+e^- \rightarrow \mbox{hadrons})/\sigma(e^+e^-\rightarrow
\mu^+\mu^-)$, where $\sigma (e^+e^- \rightarrow \mu^+\mu^-)
= \sigma^0_{\mu \mu}=4\pi \alpha^2(0) / 3s$.
%$R$ value, which counts directly the color and flavor of the quarks,
%is one of the most fundamental parameters in particle physics.
Presently the uncertainty in $R$ in the energy region below 5 GeV dominates 
the uncertainties in both $\alpha(M^2_{Z})$, the QED running coupling 
constant evaluated at the Z pole, and $a_{\mu}^{SM}$, the value of 
$(g-2)_{\mu}$ based on the Standard Model calculation~\cite{rreview_zg}.

Hadron production from $e^+e^-$ annihilation is one of the most
valuable testing grounds for Quantum Chromodynamics (QCD).
Particularly, it is interesting and important to analyze low energy
$e^+e^-$ collision data, for example, the inclusive momentum spectrum,
defined as $\xi = - \ln (2p/\sqrt{s})$, where $p$ and $\sqrt{s}$ are
the momentum of the charged particles and center-of-mass (c.m.) energy
respectively, with today's knowledge.  A purely analytical approach
giving quantitative predictions for $\xi$ is the QCD calculation using the
so-called Modified Leading Logarithmic Approximation
(MLLA)~\cite{mlla} under the assumption of Local Parton Hadron Duality
(LPHD) ~\cite{lphd}, which expresses the limiting spectrum for hadrons as

\begin{eqnarray}
\frac{1}{\sigma_{had}} \frac{d \sigma}{d \xi} = K_{LPHD} \times
        f_{MLLA} (\xi,\Lambda_{eff})
\label{express}
\end{eqnarray}

\noindent  where $K_{LPHD}$ is an overall normalization factor
describing hadronization and $f$ is a complex function of $\xi$
and effective scale parameter $\Lambda_{eff}$~\cite{mlla}. 
Eq. 1. is valid in the range of $0 \leq \xi \leq \ln
(0.5\sqrt{s} / \Lambda_{eff})$.

%Because the terms of infrared and collinear divergences can be
%factorized out, the energy dependence of the peak position of
%inclusive momentum spectra can be directly compared at parton
%level with that predicted by MLLA under LPHD assumption.

Another example is the second binomial moments, which is a measure of the
strength of hadron-hadron correlations and a sensitive probe for
higher order QCD or non-perturbative effect. It is defined as $R_2
= {\langle n_{ch}(n_{ch}-1) \rangle}/{\langle n_{ch} \rangle}^2$,
where $n_{ch}$ is the charged particle multiplicity. According to
the next leading order QCD calculation (NLO), $R_2$ is given by

\begin{eqnarray}
R_2 = \frac{11}{8}(1 - c \sqrt{\alpha_s(\sqrt{s})})
\label{r2exp}
\end{eqnarray}

\noindent with $c = 0.55 (0.56)$ for five (three) active flavors.
There has been a long standing discrepancy between the value of $R_2$ 
calculated by NLO and that measured with $e^+e^-$, $\mu^+p$ and 
$\nu_{\mu}p$ experiments. In addition, there is relatively little data
in the energy region below 5 GeV to compare with QCD calculations.

This paper presents the final results of the $R$ values measured 
with BESII~\cite{bes2} at BEPC in the energy region from 2-5 GeV. 
We also report preliminary results on the inclusive momentum spectra,
the momentum distribution of charged particles, and second binomial moments 
$R_2$ obtained from the analysis of $R$ scan data.

\section{$R$ values in 2-5 GeV}

Refs.~\cite{besr1,besr2} describe in detail the $R$ scan performed by
BES at 91 energy points between 2-5 GeV, and the experimental 
study of the background, particularly the beam associated background. 
The triggers and the determination of the trigger efficiency; the
measurement of luminosity; the hadronic event selection and
background subtraction; the determination of the detection efficiency
for hadronic events; and the initial state radiative correction
can be also found in Refs.~\cite{besr1,besr2}.

\begin{figure}[htb] 
\includegraphics[width=16pc,height=16pc]{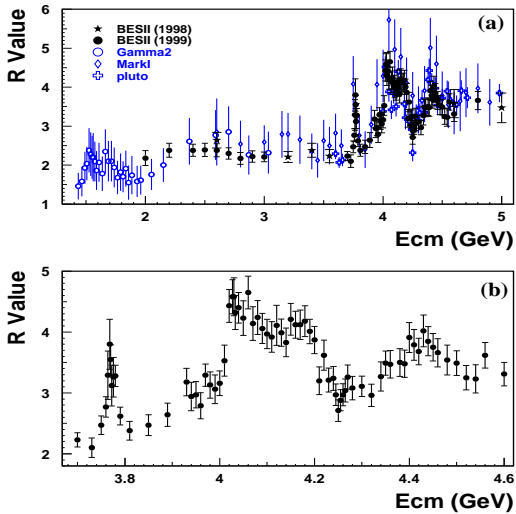}
\vskip -0.8cm
\caption{(a) A compilation of measurements of $R$ in the cm
energy range from 1.4 to 5 GeV. (b) $R$ values from this experiment 
in the resonance region between 3.75 and 4.6 GeV.} 
\label{fig:besr}
\vskip -0.4 cm
\end{figure}

The final $R$ values measured by BES in this
experiment are displayed in Fig.~\ref{fig:besr}, together with
those measured by MarkI~\cite{markI}, $\gamma\gamma 2$~\cite{gamma2} 
and Pluto~\cite{pluto}. The $R$ values from BESII have an average 
uncertainty of about 6.6\%, which represents a factor of two to 
three improvement in precision in the 2 to 5 GeV energy region.  
These improved measurements have a significant impact on the global 
fit to the electroweak data and the determination of the SM prediction 
for the mass of the Higgs particle~\cite{bolek}.  In addition, 
they are expected to provide an improvement in the precision of 
the calculated value of $a_{\mu}^{SM}$, and test the QCD sum rules 
down to 2 GeV~\cite{dave,martin,kuehn}.

\section{Test of QCD models with $R$ scan data}
 
\subsection{$\xi$ spectrum}

\begin{figure}[htbp]
\includegraphics[width=16pc,height=15pc]{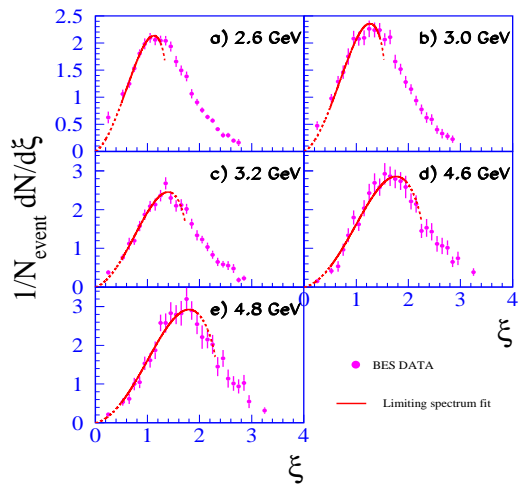}
\vskip -0.8cm 
\caption{Measured $\xi$ spectrum (solid dot) at 2.6,
3.0, 3.2, 4.6 and 4.8 GeV.  Solid curves are the fitting of the
limiting spectrum. The dotted line is an
extrapolation of the fitted result.} 
\label{comxi}
\vskip -0.4 cm
\end{figure}

The measured $\xi$ spectrum at five different energies between 2.6 and
4.8 GeV are shown in Figure~\ref{comxi}. The errors in the spectrum
include errors from hadronic event selection and uncertainties from
the event generators.

\subsection{Momentum spectrum}

\begin{figure}[htbp]
\includegraphics[width=16pc,height=15pc]{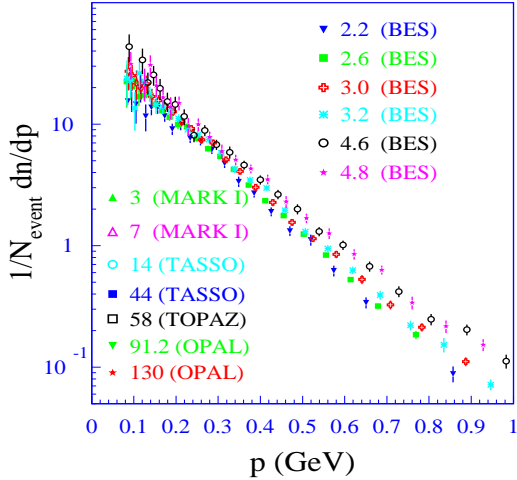}
\vskip -0.8cm 
\caption{Charged particle momentum spectrum}
\label{dnall}
\vskip -0.4 cm
\end{figure}

The momentum spectra at the five energy points are shown in
Figure~\ref{dnall}, together with those measured at higher energy
up to 130 GeV in other experiments. The momentum spectra
show that hadron production at very small momentum $p \leq 0.1$
GeV is approximately energy independent. This behavior has been
explained in Ref.~\cite{q0momentum} to be due to the coherent
emission of low energetic (i. e. long wavelength) gluons by the
total color current. Correspondingly the number of produced
hadrons at small momentum is approximately constant.

\subsection{Second binomial moment}

\begin{figure}[htbp]
\includegraphics[width=16pc,height=16pc]{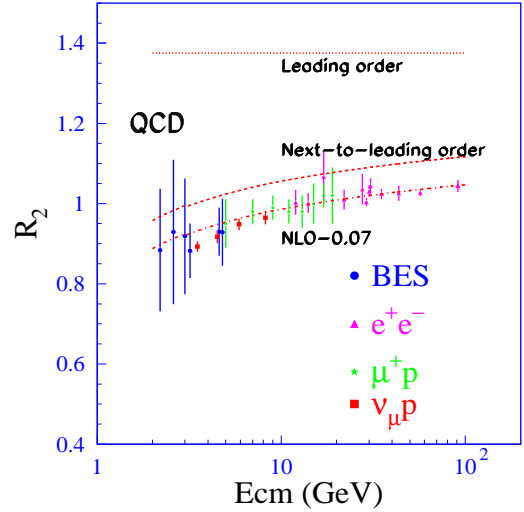}
\vskip -0.8cm 
\caption{Energy dependence of second binomial moments $R_2$} 
\label{r2bes}
\end{figure}

Based on the multiplicity measured, we can obtain the second
binomial moment $R_2$. The results are displayed in Figure
\ref{r2bes}, together with both NLO calculations and published
data at higher energies up to 100 GeV from $e^+e^-$, $\mu^+p$ and
$\nu_{\mu}p$ experiments~\cite{emcr2,wa21r2}.
Our measured $R_2$, although with large errors, are consistent with
those of other measurements done at higher energies.

It's interesting to see that $R_2$ predicted by LO QCD are
significantly higher than the measured data, while the NLO calculation
comes closer to the data, although disagreement ($\sim 0.07$ in
$R_2$) remains sizeable. 

\end{document}